# On the Error-correcting Capability of Twisted Centralizer Codes Obtained from a Fixed Rank-1 Matrix

John Ben S. Temones

Central Bicol State University of Agriculture

*johnben.temones@cbsua.edu.ph*

*Abstract* – *Twisted centralizer codes are linear codes obtained by fixing a square matrix A and collecting all square matrices B satisfying the equality AB=aBA, where a is a fixed arbitrary constant. We denote the collection of such B's as C(A,a), read as the centralizer of A, twisted by a; with every individual B's as its codewords. In this paper, we give a generalization on the error-correcting capability of twisted centralizer codes obtained from a fixed rank-1 matrix. In particular, we fix the combinatorial matrix which is obtained by getting the linear combination of square 1's matrix and identity matrix of order n. Results reveal that such codes have a dimension 1 for any fixed combinatorial matrix and constant a, hence having a relatively low information rate due to the way its codewords are constructed. However, the same family of codes are found to be maximum distance separable (MDS) codes, and hence can detect up to $n^2 - 1$ errors and can correct up to floor$\left(\frac{n^2-1}{2}\right)$ errors, both the largest possible.*

## INTRODUCTION

In 2014, a new set of linear codes known as the *centralizer codes* was introduced in [1]. Some generalizations have been made in [2] on the centralizer codes and thus gave birth to *twisted centralizer codes*. These are linear codes obtained by fixing a square matrix $A$ and collecting all square matrices $B$ satisfying the equality $AB = aBA$, where $a$ is a fixed arbitrary constant. We denote the collection of such $B$'s as $C(A, a)$, read as the centralizer of $A$, twisted by $a$; with every individual $B$'s as its codewords.

One of the results in [2] focuses on the error-correcting capability of twisted centralizer codes obtained by fixing a rank-1 matrix known as the combinatorial matrix. These matrices are of the form $xJ_n + yI_n$, where $J_n$ is the square matrix whose entries are all equal to 1 and $I_n$ is the identity matrix of order $n$ (see [4]). The result in [2] specifically suggests that twisted centralizer codes obtained from a combinatorial matrix are good error-correcting codes and can correct up to $floor\left(\frac{n^2-1}{2}\right)$ errors. However, this result is restricted to a combinatorial matrix with $x = y = 1$ and is therefore not conclusive to any ordered pair $(x, y)$.

This study is dedicated to give a generalization on the error-correcting capability of twisted centralizer codes obtained by fixing a rank-1 matrix and give implications to its probable uses in communication systems in the technological world.

## SPECTRAL PROPERTY OF COMBINATORIAL MATRIX

Combinatorial matrices, as discussed in [2], are square matrices of the form $A := xJ_n + yI_n$ where $J_n$ is the square matrix of order n whose entries are all 1 while $I_n$ is the identity matrix of order n. A combinatorial matrix is a symmetric matrix – that is, its matrix transpose is itself. It is a common knowledge as in [6] that all symmetric matrices are diagonalizable, and so are combinatorial matrices. It was further discussed in [7] that diagonalizability of a matrix is intertwined to its spectral properties.

$$A := xJ_n + yI_n = \begin{bmatrix} x+y & x & \cdots & x \\ x & x+y & \cdots & x \\ \vdots & \vdots & \ddots & \vdots \\ x & x & \cdots & x+y \end{bmatrix}$$

Fig. 1. A combinatorial matrix.

It was theorized in [1] that for a combinatorial matrix with $x = y = 1$, the only possible eigenvalues is given by the set $S := \{0, 1\}$. This study further improved this particular result in [1] by noticing that for any combinatorial matrix $A = xJ_n + yI_n$ for all ordered pairs $(x, y)$, the only possible eigenvalues for $A$ are $xn + y$ and $y$, with geometric multiplicities of 1 and $n-1$ respectively.

The proof for this theorem was obtained by particularly applying series of row and column operations on matrix $A = xJ_n + yI_n$, a method which would give us it's determinant. The $n-1$ row operations and $n-1$ column operations done in matrix $A$ could be written as a product of elementary matrices respectively. Denote by $R_i$ the elementary matrix corresponding to subtracting row 1 to the $i-th$ row of $A$ and $C_j$ the elementary matrix corresponding to adding column 1 to $j-th$ column of $A$. It can be noticed that $R_i^{-1} = C_j$ whenever $i = j$. Hence, letting a diagonal matrix $D := diag(xn+y, y, y, \ldots, y)$ with $n-1$ $y$'s, the row and column operations done to $A$ could mathematically expressed as $PAP^{-1} = D$, which could further be expressed as $A = P^{-1}DP$ where matrix $P$ is the matrix product $(R_{n-1})(R_{n-2})\ldots(R_2)(R_1)$. We found an invertible matrix $P$ hence as in [7], the only possible eigenvalues for $A$ is given by the set $S_0 := \{xn+y, y\}$, with multiplicities mentioned above.

Alternatively, this spectral property of combinatorial matrices could alternatively be proven by taking a column vector of length whose entries are all 1, denoted as the vector $u$. Notice that $Au = (xn+y)u$. The eigenspace (see [9]) of $A$ with respect to $y$ is equivalent to the null space of the matrix $xJ_n$ hence we are forced to only have two eigenvalues for $A$, which are mentioned above. This result has some implications on the error-correcting capability of $C(A,a)$ as a linear code, which is discussed in the next subsection.

## THE ERROR-CORRECTING CAPABILITY OF C(A, a) AND ITS IMPLICATIONS

The result above allows us to give a generalization for a result in [2] regarding the error-correcting capability of $C(A, a)$, which is only restricted to combinatorial matrices where $x = y = 1$. It was found out in this study that fixing any combinatorial matrix, for all ordered pairs $(x, y)$, shall give us a code with parameters $[n^2, 1, n^2]$ whenever the characteristic of the field divides $xn+y$ and $a \neq 0, 1$. The proof is detailed on the succeeding paragraphs.

Let the characteristic of the field be a prime number $p$. Since $p | xn+y$, then $A = xJ_n + yI_n$ must have an eigenvalue given by the set $S := \{0, y\}$, with multiplicities 1 and $n-1$ respectively. Hence by the previous result, $A$ must be equivalent to a certain diagonal matrix $D := (0, y, y, \ldots, y)$ with $n-1$ $y$'s. Let $E_{11}$ be the $n \times n$ matrix whose (1,1)-entry is 1 and all other entries are 0. It can be seen that $D = yI_n - yE_{11}$.

Let $B$ be a matrix in $C(D, a)$. Hence, $DB = aBD$ and is further written as $y(1-a)B = yE_{11}B - ayBE_{11}$. We write matrices $E_{11}$ and $B$ as matrices of column and row vectors, specifically,

$$E_{11} = \begin{bmatrix} & 0 & 0 & \cdots & 0 \\ & 0 & 0 & \cdots & 0 \\ e_1 & \vdots & & \ddots & \vdots \\ & 0 & \cdots & \ddots & 0 \\ & 0 & 0 & \cdots & 0 \end{bmatrix} = \begin{bmatrix} - & - & e_1 & - & - \\ 0 & 0 & \cdots & 0 & 0 \\ \vdots & \vdots & \ddots & \vdots & \vdots \\ 0 & 0 & \cdots & \ddots & 0 \\ 0 & 0 & \cdots & 0 & 0 \end{bmatrix}$$

and

$$B = \begin{bmatrix} v_1 & v_2 & \cdots & v_{n-1} & v_n \end{bmatrix} = \begin{bmatrix} u_1 \\ u_2 \\ \vdots \\ u_{n-1} \\ u_n \end{bmatrix}$$

where $e_1$ is the vector of length $n$ whose first entry is 1 and 0 elsewhere, while $v_i, u_i$ are column vectors and row vectors of length $n$, respectively, for $i = 1, 2, \ldots, n-1, n$. Hence, it can be seen that

$$E_{11}B = \begin{bmatrix} - & - & u_1 & - & - \\ 0 & 0 & \cdots & 0 & 0 \\ \vdots & \vdots & \ddots & \vdots & \vdots \\ 0 & 0 & \cdots & \ddots & 0 \\ 0 & 0 & \cdots & 0 & 0 \end{bmatrix}$$

and

$$BE_{11} = \begin{bmatrix} \hat{} & 0 & 0 & \cdots & 0 \\ \hat{} & 0 & 0 & \cdots & 0 \\ v_1 & \vdots & \ddots & \cdots & \vdots \\ \hat{} & 0 & \cdots & \ddots & 0 \\ \hat{} & 0 & 0 & \cdots & 0 \end{bmatrix}.$$

We compare the entries of $y(1-a)B$ and $yE_{11}B - ayBE_{11}$. Recall that we have the assumption that $a \neq 0,1$. Let $T = yE_{11}B - ayBE_{11}$. Let $t_{ij}, b_{ij}$ denote the $(i,j)$-th entry of $T$ and $B$, respectively. Note that the first entries of $v_1$ and $u_1$ must be the same, and is equivalent to $b_{11}$. Hence it can be seen that $t_{11}$ always agrees with $y(1-a)b_{11}$. Let $(u_1)_{1j}$ denote $j$-th entry of row vector $u_1$, $j = 1, 2, ..., n-1, n$. For $j > 1$, we have $t_{1j} = y(u_1)_{1j} = yb_{1j}$. Thus, $t_{1j} = y(1-a)b_{1j}$ if and only if $b_{ij} = 0$. Hence, $u_1$ must be a multiple of $e_1$. Let $(v_1)_{i1}$ denote the $i$-th entry of column vector $v_1$, $i = 1, 2, ..., n-1, n$. For $i > 1$, we have $t_{i1} = -ay(v_1)_{i1} = -ayb_{i1}$ and hence $t_{i1} = y(1-a)b_{i1}$ if and only if $b_{i1} = 0$. Thus $v_1$ must also be a multiple of $e_1$. Of course, for $i, j > 1$, we have $t_{ij} = 0 = b_{ij}$. Therefore, $B$ must be a multiple of $E_{11}$, i.e., $C(D,a) = \langle E_{11} \rangle$. Thus, $dim(C(D,a)) = 1$. Note that $A$ and $D$ are similar matrices thus for every $B \in C(D,a)$,

$$DB - aBD = 0_{n \times n}$$
$$(P^{-1}AP)B - aB(P^{-1}AP) = 0_{n \times n}$$
$$P^{-1}[A(PBP^{-1}) - a(PBP^{-1})A]P = 0_{n \times n}$$
$$A(PBP^{-1}) - a(PBP^{-1})A = 0_{n \times n}$$
$$A(PBP^{-1}) = a(PBP^{-1})A.$$

Hence, every element in $C(A,a)$ are the conjugates of each $B \in C(D,a)$ and therefore, $C(A,a)$ must also be 1-dimensional. It can further be noticed that

$$J_n A = AJ_n$$
$$= xnJ_n + yJ_n$$
$$= (xn + y)J_n = 0_{n \times n}.$$

Thus, $C(A,a) = \langle J_n \rangle$. Therefore, every non-zero element of $C(A,a)$ is of the form $bJ_n$ for $b \neq 0$. Hence, $C(A,a)$ has minimum distance equal to $n^2$. The length of each codeword in $C(A,a)$ must be noted to be $n^2$ due to the way codewords are constructed in [1] and [2]. Therefore, $C(A,a)$ has parameters $[n^2, 1, n^2]$, as desired.

The result above guarantees us to always have a maximum distance separable (MDS) codes for twisted centralizer codes obtained from combinatorial matrices conforming to the conditions set by the theorem above. MDS codes in [9] are known to attain the Singleton bound hence are good error-correcting codes. In particular, such twisted centralizer code could detect up to $floor\left(\frac{n^2-1}{2}\right)$ errors, also the highest possible. This is good, however, the result above also implies that the same twisted centralizer code will have a relatively low information rate which is found to be only at $\frac{dimension\ of\ code}{length\ of\ code} = \frac{1}{n^2}$.

The main linear coding theory (MLTC) problem is about. It was discussed in [4] that a good (linear) code should have a relatively low codeword length as it lowers the rate to which information are being sent through a noisy communication channel. Vis-a-vis, in order to have a good information rate, a code must have a relatively higher code dimension. The dimension of the code also gives us the leeway to code more messages in our system, hence a larger one is also recommended. On the other end, an attribute of a good code is if it can detect or even correct a large amount of errors whenever faulty messages have been sent to a receiver. The error-correcting capacity of a code with minimum distance $d$ is defined by the expression $floor\left(\frac{d-1}{2}\right)$, and hence a good error-correcting code should also have a relatively higher minimum distance. Optimizing the code is the problem, as too much of one side compromises that of the other parameter(s).

The result above is a perfect example of what MLTC problem talks about. It guarantees us to always have a maximum distance separable (MDS) codes for twisted centralizer codes obtained from combinatorial matrices conforming to the conditions set by the theorem above. MDS codes in [9] are known to attain the Singleton bound hence are good error-correcting codes. In particular, such twisted centralizer code could detect up to $floor\left(\frac{n^2-1}{2}\right)$ errors, also the highest possible. This is nice, however, the result above also implies that the same twisted centralizer code will have a relatively low information rate which is found to be only at $\frac{1}{n^2}$. This means that for relatively large $n$, we can have a code which could be relied to give us the original message when it gets distorted as it is travelling through the communication channel, but we are also sacrificing the rate to which this information are being sent. Low dimension (in this case, the code always gives us dimension 1) means we can also code a relatively small number of messages in our communication system.

## CONCLUSION AND RECOMMENDATION

This paper was an attempt to generalize a result in [2] on the error-correcting capability of twisted centralizer codes obtained from combinatorial matrix; a rank-1 matrix obtained from the linear combination of the square matrix with entries all equal to 1 and the identity matrix. Results revealed that the only possible eigenvalues for a combinatorial matrix $A = xJ_n + yI_n$ is given by the set $S \coloneqq \{xn + y, y\}$, with multiplicities equal to 1 and $n - 1$ respectively. The spectral property of combinatorial matrix plays a role on the error-correcting capability of twisted centralizer code with fixed matrix $A$, as it was found out that an MDS code is guaranteed to be obtained, with parameters $[n^2, 1, n^2]$. The code obtained poses a highly accurate and reliable error-detecting and error-correcting capacity but is limited with coded messages and has low information rate. Future researchers in this area may consider finding ways or condition to optimize the three parameters of a twisted centralizer code. Moreover, one may consider fixing a matrix with rank higher than 1 and study the error-correcting capability of the twisted centralizer code.